\documentstyle[aps,preprint,eqsecnum]{revtex}

\def \be {\begin{equation}}
\def \eq {\end{equation}}
\def \bee {\begin{eqnarray}}
\def \eqq {\end{eqnarray}}
\def \nn {\nonumber}
\def \bea {\begin{array}{c}}
\def \eqa {\end{array}}

\def \del {\partial}
\def \dels {\partial\kern-.5em / \kern.5em}
\def \As {{A\kern-.5em / \kern.5em}}
\def \Ds {D\kern-.7em / \kern.5em}

\def \H {{\cal H}}

\def \xih {\hat{\xi}}

\def \a {\alpha}
\def \b {\beta}

\def \m {\mu}
\def \n {\nu}

\def \O {\Omega}

\begin{document}
\draft
\title{Near-Horizon Virasoro Symmetry and \\
the Entropy of de Sitter Space in Any Dimension}
\author{Feng-Li Lin and Yong-Shi Wu}
\address{Department of Physics, University of Utah, 
Salt Lake City, UT 84112, U.S.A.}
\date{\today}
\maketitle
\begin{abstract}
De Sitter spacetime is known to have a cosmological 
horizon that enjoys thermodynamic-like properties 
similar to those of a black hole horizon. In this 
note we show that a universal argument can be given 
for the entropy of de Sitter spacetime in arbitrary 
dimensions, by generalizing a recent near horizon 
symmetry plus conformal field theory argument of 
Carlip for black hole entropy. The implications of 
this argument are also discussed.
\end{abstract}

\newpage

\section{Introduction}

In general relativity, gravitational collapse of massive
stars and the expansion of the universe are two sides of 
the same coin. Both are manifestation of the instability 
of gravitation. The singularity theorem of Hawking and 
Penrose \cite{HP} applies to both. In its final stage, 
gravitational collapse always leads to formation of an 
event horizon, the black hole horizon, which is the 
boundary of a spacetime region which is not visible to
an external observer. In cosmology the counterpart of
the black hole horizon is the cosmological event 
horizon\footnote {Hereafter we will use simply the 
term ``horizon'' in lieu of ``event horizon''.} in 
a de Sitter universe. 

The de Sitter metric in $d$ dimensions is 
given by
\be
ds^2= - (1-\frac{r^2}{\ell^2}) dt^2 + 
(1-\frac{r^2}{\ell^2}) dr^2 + r^2 d\O^2,
\label{dS}
\eq
where $d\O^2$ is the solid-angle element on 
$(d-2)$-dimensional sphere, and the range 
$0\le r \le \ell$ covers a portion of de Sitter 
space with the boundary at $r=\ell$. With 
$d=4$, it was first discovered \cite{deS} as 
a vacuum solution to the Einstein equations with 
a replusive cosmological constant 
$\Lambda= (d-1)(d-2)/ 2 \ell^2$. On one hand, 
cosmological models with a repulsive cosmological 
constant which expands forever approach 
asymptotically de Sitter space at large times. 
On the other hand, the exponential expansion 
of our universe in the early inflationary 
period was driven by the vacuum energy of a 
scalar field, effectively acting as a positive 
cosmological constant, and thus can be described 
by the de Sitter metric. 

It is well-known that a de Sitter universe expands 
so rapidly that for the geodesic observer at the 
origin, there is an cosmological horizon located 
at $r=\ell$, from beyond which light can 
never reach him/her. The area of this horizon may 
be regarded as a measure of his/her lack of 
knowledge about the rest of the universe beyond 
his/her ken. Thus, one expects that a cosmological 
horizon should have many similarities with a black 
hole horizon. Indeed in late seventies Gibbons and 
Hawking \cite{GH} have shown that in general 
relativity, both the black-hole and the cosmological 
horizons share the same set of laws which are 
formally analogous to those of thermodynamics: In 
either case, the surface gravity at the horizon is 
proportional to the effective temperature, and 
the area $A$ of the horizon is to the entropy 
$S$, as given by the Bekenstein-Hawking formula:
\be
S={A \over 4G}\;.
\label{area}
\eq
Furthermore, they 
showed that if the quantum effects of pair creation 
in curved spacetime are included, this similarity 
between the laws of horizons and thermodynamics is 
more than an analogy: An observer will detect a 
background of thermal radiation coming apparently 
from the cosmological horizon, in a manner similar 
to the Hawking radiation from a black hole horizon. 
Thus the close connection between horizons and 
thermodynamics has a wider validity than the 
ordinary black hole cases in which it was first 
discovered\cite{GH}. 

This lesson becomes particularly important in the 
wake of the recent progress in string theory in 
understanding the microscopic states that are 
responsible for the black hole entropy \cite{SV,BHRev}. 
One is naturally led to search for a fundamental, 
microscopic mechanism for gravitaional entropy that
is {\it universally applicable to de Sitter entropy 
in arbitrary dimensions as well as to black hole 
entropy.} The method of counting D-brane 
states for black hole entropy does not seem to 
satisfy this universality requirement, since at 
present we do not know yet what are the D-brane 
states responsible for a cosmological horizon. 
Recently Maldacena and Strominger \cite{MS} has 
attacked the problem of de Sitter entropy in the 
particular case of $2+1$ dimensions. They
explored the equivalence between Chern-Simons 
gauge theory and $2+1$ de Sitter gravity, and
showed that the asymptotic symmetry group of 
the theory near the cosmological horizon 
contains a Virasoro subalgebra\cite{Park}, 
with a central 
charge right to reproduce the de Sitter entropy 
via the Cardy formula \cite{Cardy,Carlip0}
\be
log (\rho(h,\bar{h}))=2\pi \left(\sqrt{ch \over 6}
+ \sqrt{c\bar{h} \over 6} \right), 
\label{Cardy's}
\eq
which counts the asymptotic density of 
states of the Hilbert space of a conformal 
field theory (CFT) labelled by the 
conformal weight $(h,\bar{h})$ and the 
(effective) central charge $c$. However, 
it is hard to see how this argument 
could be generalized to quantum gravity 
in other dimensions, where no equivalent 
Chern-Simons theory is available.

Recently Carlip \cite{Carlip} has put forward
a universal argument for black hole entropy 
in any dimension, exploring near-horizon
symmetry and conformal field theory. In
the present note we will show that his 
argument can be generalized to give a 
universal argument for the de Sitter 
entropy in arbitrary dimensions. Also a 
recent proposal in ref. \cite{Solodukhin} 
for a candidate CFT on black horizon by 
dimensional reduction of gravity will 
be shown to be applicable to the de 
Sitter case. This note not only adds 
weight to the universality of Carlip's 
argument, but also points to a profound 
connection between the microscopic 
origin of the gravitational entropy 
associated with any horizon and conformal 
field theory. Indeed for quite a while 
there have been suggestions on the close
relationship between the black hole 
entropy and conformal field theory, see
e.g. \cite{LW} and references therein. 
Our note confirms the existence of such 
a relationship for cosmological horizons. 

\section{Near-Horizon Virasoro Symmetry} 

Let us consider a ``sector'' in quantum gravity
that, in the semiclassical limit, corresponds to
fluctualtions (of both geometry and coordinates) 
around the standard de Sitter metric (\ref{dS}): 
\begin{equation}
ds^2= -N^2 dt^2 + f^2 (dr+ N^r dt)^2 \
+\sigma_{\alpha\beta} (dx^\alpha+N^\alpha dt) 
(dx^\beta+N^\beta dt). 
\label{ddS} 
\end{equation}
Here $x^{\alpha}$ are coordinates on a 
$(d-2)$-dimensional sphere. We have adopted 
the Arnowitt-Deser-Misner scheme, fixing 
the horizon to be located at $r= \ell$ 
and requiring that near the horizon the lapse 
function $N$ behave as 
\be
N^2= \frac{4\pi}{\b} (\ell-r) + O (\ell -r)^2,
\label{lapse}
\eq
where $\b\equiv 2\pi \ell$. We will treat the 
horizon as an outer boundary (i.e. 
$0\le r \le \ell$) and require that the 
metric approach that of a standard de Sitter 
metric on this boundary. To define the theory 
more precisely, one may impose the following 
fall-off conditions near the boundary:
\bee
f=O(N^{-1}), \;\; N^r=O(N^2),\;\;
\sigma_{\a\b}=O(1), N^\a=O(N),\nn \\
(\partial_t - N^r \partial_r) g_{\m\n}
= O(N) g_{\m\n}, \;\; 
\nabla_\a N_\b+ \nabla_\a N_\a
=O(N^2).
\label{falloff}
\eqq
similar to those in ref. \cite{Carlip},
with the only difference in the equations
involving $N_\a$, which in our case
essentially require that the angular 
momentum be constantly vanishing on the 
horizon. Given this asymptotic behavior of 
the metric, it is easy to check that near 
the horizon, the extrinsic curvature of a 
slice of constant time behaves like
\be
K_{rr}= O(N^{-3}), \;\; K_{\a r}= O(N^{-2}),
\;\; K_{\a\b}= O(1).
\eq

In the same spirit of the work of Brown and 
Henneaux \cite{BH} and of Carlip \cite{Carlip}, 
we want to show that the gauge symmetries of 
this classical theory with boundary contains 
a Virasoro subalgebra with a central charge. 
In the Hamiltonian formulation of general 
relativity, the gauge symmetries are the 
so-called surface deformations\footnote{
Readers who are unfamiliar with the 
formulation of the surface deformations can 
see \cite{Teitelboim} for details.}  that 
preserve the fall-off conditions (\ref{falloff})
near the horizon. The full generator for surface 
deformations is known to be given by
\be
L [\xih]=H [\xih]+ J [\xih],
\label{fgene}
\eq
where the first term is the bulk term and 
the second the boundary term \cite{BH}:
\bee
H [\xih]&=&\int_\Sigma d^{d-1}x \; \xih^\mu 
\;  \H_\mu\,  
\nonumber \\
J [\xih] &=& \frac{1}{8\pi G} 
\int_{r=\ell} d^{d-2}x
\left\{ n^a \nabla_a \xih^t \sqrt{\sigma}
+ \xih^a \pi_a^r + n_a \xih^a K 
\sqrt{\sigma}\right \}.
\label{gene}
\eqq
Here $\{\H_t, \H_a \}$ $(a=r, \a)$
are the Hamiltonian and momentum 
constraints. The surface deformation 
parameters $\xih^\mu$ are related 
to spacetime diffeomorphism 
parameters $\xi^\mu$ by 
$\xih^t=N\xi^t$ and $\xih^a= 
\xi^a + N^a \xi^t$. 
To preserve the fall-off 
conditions (\ref{falloff})
it is required that near $N=0$,
\be 
\xih^t = O(N),\;\; \xih^r= O(N^2), \;\;
{\rm and} \;\; \xih^\a = O(1).
\eq
They satisfy a Lie algebra; the part 
relevant to our later analysis is
\be
\{\xih_m,\xih_n\}^t_{SD}=
\xih^a_1\partial_a\xih^t_2 
-\xih^a_2\partial_a\xih^t_1.
\label{Lie}
\eq

In the full generator (\ref{fgene}), we need 
the boundary term $J [{\hat{\xi}}]$, whose 
variation cancels the unwanted surface term 
in the variation of the bulk term $H [\xih]$, 
so that the functional derivative of $L[\xih]$ 
is well defined. It is easy to verify that 
this is indeed true if we restrict our 
variations to those satisfying
\be 
\delta f/f = O(N),\;\; 
\delta K_{rr}/K_{rr}= O(N).
\eq

Following \cite{Carlip}, we consider a 
particular class of surface deformations 
as follows:  

First, for simplicity, we specialize 
to the cases with $0=N^r=N^\a=
\partial_{t,\phi}(N^{\phi})=
\partial_{t,\phi}(N^2)$, ($\a\neq\phi$),
where $\phi$ is a selected azimuthal 
angle such that $N^{\phi}$ is $O(N)$ 
but not zero. Later we will see that
this will help us to distinguish 
the left and right circular modes of 
$\xi^{\mu}$. To avoid the singular 
behavior on the horizon for the 
angular mode-decomposition of $\xi^t$,
we introduce a spherical surface 
$H_{\epsilon}$ with distance 
$\epsilon$ to the horizon, and
take $N^{\phi}$ to be constant
on this surface, which tends to 
zero as we finally take $\epsilon 
\rightarrow 0$ at the end of the 
calculation.     

Second we single out a particular
class of surface deformations satisfying
the $Diff(S^1)$ algebra by the following
conditions:
 
\begin{itemize} 
\item{1.} In the conformal coordinates 
defined by $fdr=Ndr^*$, the red-shift 
effect makes the diffeomorphism 
$\xi^t$ to be light-like, and the 
classical nature of the horizon allows 
only the outgoing one into the horizon, 
that is $(\del_t+\del_{r^*})\xi^t=0$.

\item{2.} We restrict to the surface 
deformations which do not change the location
of the horizon defined by the zero of the
lapse function $N^2$. We then impose 
$\delta_{\xi^{\mu}} g^{tt}=0$, where $g^{tt} 
\equiv{-1 \over N^2}$. This leads to 
$(\del_t-N^{\phi}\del_{\phi})\xi^t
+[(N^2)_{,r}/ 2 N^2] \xi^r=0$ 
on $H_{\epsilon}$.
Then the right circular modes defined by  
$(\del_t-N^{\phi}\del_{\phi})\xi_{(+)}=0$ 
becomes zero, and the components of the
left circular modes defined by
$(\del_t+N^{\phi}\del_{\phi})\xi_{(-)}=0$ 
satisfy the relation 
\be
\xi_{(-)}^r={-2N^2 \over (N^2)_{,r}}
(\del_t-N^{\phi} \del_{\phi})\xi_{(-)}^t
={-4N^2 \over (N^2)_{,r}}
\del_t \xi_{(-)}^t\;.
\eq
 
\item{3.} If we assume the parameter 
$\xi_{(-)}^t$ 
to be periodic in time, with period 
$T$ when analytically continued to the 
Euclidean signature, then from (1) and (2) it 
can be decomposed into the Fourier modes 
\be
\xi^t_{(-)n}=a_n exp\{\frac{2\pi in}{T}
(t-r^*-{\phi \over N^{\phi}})\}.
\label{modes}
\eq
Note that the angular decomposition would
become singular on the horizon since
$N^{\phi}=0$ there; however, in our 
treatemnt this is avoided by evaluating 
everything first on $H_{\epsilon}$,
not directly on the horizon.
 
\item{4.} From (1), (2) and (3), the
surface deformaion Lie algebra (\ref{Lie})
becomes
\be
\{\xih_{(-)m},\xih_{(-)n}\}^t_{SD}
=i(n-m) \xih^t_{(-)m+n} + 
\xih_{m}^{\phi}\del_{\phi}\xih_{(-)n}^t
-\xih_{n}^{\phi}\del_{\phi}\xih_{(-)m}^t\;,
\label{DA}
\eq
if $a_n = {T / 4\pi}$. Clearly, (\ref{DA}) 
reduces to $Diff(S^1)$ only if we impose 
$\xih^{\phi}=0$, which implies that we 
restrict the surface deformations to the
$r-t$ plane. We emphasize (\ref{DA}) is 
exact without using any fall-off condition, 
so it is valid away from the horizon.
\end{itemize}

To show that the above class 
of surface deformations generate a 
Virasoro algebra, and to calculate 
its central charge, we
invoke the well-known 
fact \cite{BH,Carlip} 
that the Poisson brackects 
of generic surface 
deformations close to the 
Lie algebra of 
surface deformations, 
$\{ \xih_1,\xih_2 \}_{SD}$, 
with a possible central term 
$K[\xih_1,\xih_2]$:
\be
\left\{ L [\xih_m],L[\xih_n] \right\}
= L [\{\xih_m, \xih_n \}_{SD}]
+ K[\xih_m, \xih_n]=i(n-m)L[\xih_{m+n}]
+K[\xih_m,\xih_n]\;.
\label{SD}
\eq
 
In ref. \cite{Carlip}, it has been shown
that the right-hand side of (\ref{SD}) is 
given by the boundary variation of 
(\ref{fgene}), (\ref{gene}), which 
in our case yields
\be
\frac{1}{8\pi G}\int_{r=\ell -\epsilon} 
d^{d-2}x \sqrt{\sigma}
\{\frac{1}{f^2}\del_r(f\xih^r_n)
\del_r\xih^t_m+\frac{1}{f}
\del_r(\xih^r_m \del_r\xih^t_n)
-(m \leftrightarrow n) \}.
\label{bv}
\eq
To have a well-defined angular 
decomposition, this integral is defined 
on the hypersurface $H_{\epsilon}$. The 
result turns out to be independent of
$\epsilon$, so it is safe to take the 
limit $\epsilon \rightarrow 0$ in the 
final result. Also,
there is an overall sign difference from 
\cite{Carlip} because of the reversed 
direction of the outward unit normal as 
compared to that of the black hole horizon. 

Now let us substitute the modes 
(\ref{modes}) into (\ref{bv}) 
and evaluate it on shell at the de 
Sitter metric on $H_{\epsilon}$, then 
(\ref{SD}) becomes
\be
\left\{ L [\xih_{(-)m}],L[\xih_{(-)n}] 
\right\}|_{H[\xih]=0} 
= i \; \frac{A}{8 \pi G} \frac{\b}
{T} n^3 \delta_{m+n,0}
=i\;(n-m)J[\xih_{(-)m+n}]+
K[\xih_{(-)m},\xih_{(-)n}],
\label{onshell}
\eq
where $A$ is the area of the 
cosmological horizon.  
From  (\ref{gene}) one obtains
\be
J[\xih_{(-)m}]=\frac{A}{16 \pi G} 
\frac{T}{\b} \delta_{m,o}\;,
\label{L0}
\eq
Thus,
\be
K[\xih_{(-)m},\xih_{(-)n}]= i \; \frac{A}
{8 \pi G} \frac{\b}{T} 
(n^3-n\; \frac{T^2}{\b^2}) \delta_{m+n,0} 
\label{K}
\eq 
If we define $L_m=L[\xih_{(-)m}]$, 
then eq. (\ref{SD}) gives us a Virasoro 
algebra with central charge\footnote{
This formula is true for any value of 
the period $T$; however, a preferred 
value is $T=\b$, with which the horizon 
is free of conical singularity.}
\be 
c=\frac{3A}{2\pi G}{\b \over T}\; .
\label{cq}
\eq

We view the above classical symmetry as 
resulting from that of the quantum theory 
of gravity. Thus, we infer that the latter 
should respect a (chiral copy of) 
Virasoro algebra with the 
central charge (\ref{cq}), and the quantum 
states characterizing the cosmological
horizon must form a representation of
this algebra with the conformal weight 
$h= (A/16\pi G) (T/\b)$, read from 
(\ref{L0}). Then one applies Cardy's 
formula (\ref{Cardy's}) to count the 
asymptotic density of states, and get 
the correct entropy (\ref{area}) for 
the cosmological horizon. \footnote{To 
get the standard form of the central 
term in (\ref{K}), one may shfit $L_0$,
and therefore $h$, by $c/24-h$; this 
also causes a shift in $c$ 
by $24h-c$, which makes eq. 
(\ref{Cardy's}) invariant.}

\section{CFT Candidate by Dimensional 
Reduction} 

In the last section we have used 
symmetry 
arguments to derive the behavior 
of any
quantum mechanical theory of 
cosmological 
horizon states. But they did not 
answer the 
question of what are the specific 
degrees of 
freedom that account for the 
horizon states.
Recently Solodukhin\cite{Solodukhin}  
has suggested 
one possible candidate for black 
hole horizon states 
by dimensional reduction.
In this section, we will show that it is 
possible to adapt his procedure to 
the case 
of a cosmological horizon, yielding a 
conformal field theory (CFT) with 
the same 
entropy we have just derived. 

Let us start with the Einstein-Hilbert action 
with a positive cosmoloical constant 
$\Lambda$:
\be 
S_{(d)}=-{1 \over 16\pi G_d}
\int_{M^d} d^dx\; 
\sqrt{-g_{(d)}}\; (R_{(d)}-2\Lambda)\;.
\eq
where $G_d$ is the Newton constant in 
d-dimensional space-time. We assume 
spherical symmetry so that the metric 
is of the following form ($a,b=0,1$)
\be 
ds^2=\gamma_{ab}\;dx^adx^b+r^2(x_0,x_1)
d\Omega^2\;,
\label{sphere}
\eq
where $\gamma_{ab}$ is the metric of 
the destined 2-dimensional manifold and 
$r^2$ represents the degrees of freedom
for spherical symmetric fluctuations of 
the $(d-2)$-dimensional spheres.
Moreover, the 2-dimensional part of 
the metric will have the following 
near horizon behavior as in the de 
Sitter background,
\be 
ds^2_{(2)}=-N^2\;dt^2+\frac{dr^2}
{N^2},
\eq
where $N^2$ is the asymptotic 
lapse function (\ref{lapse}).

After dimension reduction with the 
metric in the form of (\ref{sphere}), 
the action is reduced to an 
effective 2-dimensional theory
\be
S_{(2)}=-\int_{M^2} d^2x\; 
\sqrt{-\gamma}\; 
\{{1 \over 2} (\del \Phi)^2 + 
{1 \over 8} 
({d-2 \over d-3}) \Phi^2 (R_{(2)}
-2\Lambda
+ ({C \over \Phi^2})^{({2 \over d-2})} 
\Omega_{(d-2)}) \}\;.
\label{action2}
\eq
Here $\Omega_{(d-2)}=(d-2)(d-3)$ 
is the scalar curvature of the 
$(d-2)$-dimensional unit sphere, and
\be
\Phi^2=C r^{d-2}\;,  \qquad C
={\Sigma_{d-2} \over 2\pi G_d}
({d-3 \over d-2})\;,
\eq
where $\Sigma_{d-2}$ is the area 
of the unit sphere $S^{d-2}$.

Apply the following substitution
\be
\gamma_{ab}\rightarrow 
({\phi_h \over \phi})^{({d-3 
\over d-2})} \; e^{{1 \over Q}\phi} 
\gamma_{ab}\;, 
\qquad \Phi^2 \rightarrow 4 
({d-3 \over d-2}) Q \phi.
\eq
where $\phi_h$ is the value of 
$\phi$ at the horizon 
$r=\ell$. Then (\ref{action2}) is 
transformed into the 
familiar Liouville-type form
\be
S_L=-\int_{M^2} d^2x\; 
\sqrt{-\gamma}\; \{{1 \over 2} 
(\del \phi)^2 + {1 \over 2} Q  
\phi R_{(2)} +U_d(\phi) \}
\label{Liouville}
\eq
where the scalar potential
\be
U_{d}(\phi)= \{ ({\Sigma_{d-2} \over 
16\pi G_d})^{2 \over d-2}
({Q\phi \over 2})^{d-4 \over d-2}
- Q \Lambda \phi) \} ({\phi_h \over 
\phi})^{({d-3 \over d-2})} \; 
e^{{1 \over Q}\phi} \;.
\eq
Note that $U_d(\phi_h)$ is finite.  
Our $U_d$ only differs from 
\cite{Solodukhin} by the term 
involved $\Lambda$,
which is irrevelant in deriving 
the Virasoro algebra below.

The trace of the stress tensor 
derived from (\ref{Liouville}) 
by varying $\gamma^{ab}$ is
\be
T\equiv \gamma^{ab} T_{ab} = 
{-1 \over 2}Q \Box \phi 
+ U_d(\phi)  
\eq

It is obvious that the theory is not  
conformal because of the nonzero $T$. 
However, near the horizon, 
the red-shift effect will suppress 
the self interactions 
$U_d(\phi_h)$\cite{Solodukhin}, 
and the theory 
will become conformal in 
the following coordinates:  
\be
z=\int {dr \over N^2(r)}=
{-\beta \over 4\pi} ln(\ell -r)\;.
\eq
One remark here is that even after 
suppressing the scalar 
potential $U_d$, $T$ will vanish 
only if the equation of motion of 
$\phi$ in the new coordinate is 
satisfied, so the CFT is purely
classical. Because the incoming motion 
from the horizon is forbidden in the 
classical theory,  one can only use 
the component $T_{++}=T_{tt}+T_{tz}$ 
of the stress tensor, not the 
component $T_{--}$, as the 
``physical'' charge generating the 
conformal transformations. 
So the resultant CFT is chiral. 
Note that this arguments for the
chiral nature of CFT is 
different from the condition (2) 
used in Sec. 2. 

Following \cite{Solodukhin}, the 
Virasoro algebra generated by $T_{++}$ 
can be shown to have the central 
charge $c=12\pi Q^2$. 
If we write $Q=q\Phi_h/2$  , 
then the central charge becomes
\be
c=6q^2 ({d-3 \over d-2}) {A \over 4 G_d}\;.
\eq
The value of $L_0$ can be determined in a
way similiar to\cite{Solodukhin} to be
\be
h={1 \over 4\pi^2 q^2}({d-2 \over d-3})
{A \over 4G_d}.
\eq
Using Cardy's formula this leads to
the same entropy (\ref{area}) for
cosmological horizon, independent of
the parameter $q$.

\section{Discussions}

Some remarks are in order.

1) We have shown that the close relationship 
between gravitational entropy associated 
with event horizon and conformal 
field theory that was found for black holes
actually has a wider validity, i.e. it holds
also for cosmological de Sitter-like horizon.

2) The central charge calculated above for
the de Sitter horizon, either in Sec. 2 or
in Sec. 3, is {\it classical in nature}. 
Namely it is the central charge in a Virasoro 
algebra in the classical theory of gravity.

3) Why the classical central charge gives 
us the correct entropy when we blindly
apply the Cardy formula in quantum conformal 
field theory is still a mystery not well
understood yet. The applicability of the 
Cardy formula for de Sitter entropy actually 
involves several key assumptions about the
conformal field theory associated with quantum 
gravity. These assumptions are essentially the
same as those involved in the conformal field 
theory arguments for black hole horizons,
as metioned and discussed in the literatures
\cite{MS,Carlip,Carlip0,Strom}. We 
would not like to repeat them here.

4) However, in view of the fact that the
value of the classical central charge
gives the correct value for the gravitational
entropy when combined with the Cardy
formula, we naturally expect that the 
value of the classical central charge would 
not get modified in quantum gravity. Perhaps 
this suggests that the correct quantum theory 
of gravity should not be a theory that 
quantizes the classical theory of gravity.
Rather the latter is a low-energy effective 
theory of the former. 

One of us, F.L.L., thanks a correspondence 
with Dr. Solodukhin. This work was supported 
in part by NSF grand PHY-9601277.

\end{document}